\documentclass[fleqn,usenatbib]{mnras}

\usepackage{hyperref}

\usepackage{subfigure} 

\usepackage[T1]{fontenc}

\DeclareRobustCommand{\VAN}[3]{#2}
\let\VANthebibliography\thebibliography
\def\thebibliography{\DeclareRobustCommand{\VAN}[3]{##3}\VANthebibliography}

\usepackage{enumitem}

\usepackage[pdftex]{graphicx}

\usepackage{amsmath}	

\usepackage{lmodern} 

\usepackage{newtxtext,newtxmath} 
\usepackage{bm}

\usepackage{hyperref} 
\usepackage{natbib}   
\defcitealias{Castor1975}{CAK}

\title[Line-driven accretion disc winds]{Monte-Carlo radiation hydrodynamic simulations of line-driven disc winds: relaxing the isothermal approximation}
    
\author[A Mosallanezhad et al.]{
Amin Mosallanezhad,$^{1}$\thanks{E-mail: a.mosallanezhad@soton.ac.uk}
Christian Knigge,$^{1}$ \thanks{E-mail: c.knigge@soton.ac.uk}
Nicolas Scepi,$^{2}$ 
James H. Matthews,$^{3}$
Knox S. Long,$^{4,5}$ \thanks{E-mail: long@stsci.edu}
\newauthor
Stuart A. Sim,$^{6}$
and Austen Wallis$^{1}$
\\
$^{1}$School of Physics and Astronomy, University of Southampton, Highfield, Southampton SO17 1BJ, UK\\
$^{2}$CNRS, IPAG, Universit'e Grenoble Alpes, F-38000 Grenoble, France\\
$^{3}$Department of Physics, Astrophysics, University of Oxford, Denys Wilkinson Building, Keble Road, Oxford OX1 3RH, UK \\
$^{4}$Space Telescope Science Institute, 3700 San Martin Drive, Baltimore, MD 21218, USA\\
$^{5}$Eureka Scientific Inc., 2542 Delmar Avenue, Suite 100, Oakland, CA 94602-3017, USA\\
$^{6}$School of Mathematics and Physics, Queen’s University Belfast, University Road, Belfast BT7 1NN, UK
}

\date{\today}

\pubyear{\the\year{}}

\begin{document}
\label{firstpage}
\pagerange{\pageref{firstpage}--\pageref{lastpage}}
\maketitle

\begin{abstract}
Disc winds play a crucial role in many accreting astrophysical systems across all scales. In accreting white dwarfs (AWDs) and active galactic nuclei (AGN), radiation pressure on spectral lines is a promising wind-driving mechanism. However, the efficiency of line driving is extremely sensitive to the ionization state of the flow, making it difficult to construct a reliable physical picture of these winds. Recently, we presented the first radiation-hydrodynamic (RHD) simulations for AWDs that incorporated detailed, multi-dimensional ionization calculations via fully frequency-dependent radiative transfer, using the \textsc{Sirocco} code coupled to \textsc{Pluto}. These simulations produced much weaker line-driven winds ($\dot{M}_{\rm wind}/\dot{M}_{\rm acc} < 10^{-5}$ for our adopted parameters) than earlier studies using more approximate treatments of ionization and radiative transfer (which yielded $\dot{M}_{\rm wind}/\dot{M}_{\rm acc} \simeq 10^{-4}$). One remaining limitation of our work was the assumption of an isothermal outflow. Here, we relax this by adopting an ideal gas equation of state and explicitly solving for the multi-dimensional temperature structure of the flow. In the AWD setting, accounting for the thermal state of the wind does not change the overall conclusions drawn from the isothermal approximation. Our new simulations confirm the line-driving efficiency problem: the predicted outflows are too highly ionized, meaning they neither create optimal driving conditions, nor reproduce the observed ultraviolet wind signatures. Possible solutions include wind clumping on sub-grid scales, a softer-than-expected spectral energy distribution or additional driving mechanisms. With the physics now built into our simulations, we are well-equipped to also explore line-driven disc winds in AGN.
\end{abstract}

\begin{keywords}
accretion, accretion discs -- hydrodynamics -- radiative transfer -- methods: numerical -- novae, cataclysmic variables -- stars: winds, outflows.
\end{keywords}


\section{Introduction}

Winds are a prominent feature of numerous astrophysical systems, particularly those involving accretion discs. Such environments, including active galactic nuclei (AGN; e.g. \citealt{Gibson2009}), X-ray binaries (XRBs; e.g. \citealt{DiazTrigo2016}; \citealt{Tetarenko2018}), and accreting white dwarfs (AWDs; e.g. \citealt{Froning2012}) frequently exhibit blueshifted absorption lines, providing clear evidence of outflowing gas.
These outflows can profoundly alter the appearance of the entire system. Wind signatures observed in UV resonance lines serve as essential tools for studying outflows in astrophysical systems such as AGN. These features arise from the absorption and re-emission of UV photons by ions such as C~IV, Si~IV, and N~V, providing insights into the velocity, ionization structure, and geometry of the winds. Moreover, the characteristic \textit{P Cygni} profiles and the broad absorption lines (BALs) observed in UV spectra reveal the presence of high velocity outflows. These outflows are commonly interpreted as being driven by radiation pressure on resonance lines--a process known as \textit{line-driven winds}. In AGN, direct evidence for the importance of line driving comes from the \textit{ghost of Ly-$\alpha$} (e.g. \citealt{Arav1995}; \citealt{Arav1996}; \citealt{MasRibasMauland2019}; \citealp{Filbert2024}) and instances of \textit{line locking} (e.g. \citealt{Korista1993}; \citealt{LuLin2018}) observed in the spectra of more distant luminous AGNs, such as quasi-stellar objects (QSOs), and BAL QSOs.

Beyond their observational signatures, outflows can play a crucial role as a mass, energy, and momentum sink for the underlying accretion system, while simultaneously acting as a source of these quantities for the surrounding environment. In some AWDs (\citealt{Scepi2019}), XRBs (e.g. \citealt{Ponti2012}; \citealt{Higginbottom2019}) and ultraluminous X-ray sources (ULXSs; \citealt{Fabrika2015}; \citealt{Middleton2014, Middleton2022}), a substantial fraction -- if not the majority -- of the accreted material may be ejected. Consequently, the rate of accretion onto the central object can be significantly lower than the rate at which material is supplied to the outer disc. The broader influence of outflows is perhaps most clearly illustrated by disc winds and jets driven from AGN and quasars. These outflows enable supermassive black holes (SMBHs) to interact with their host galaxies and clusters on large scales, providing a crucial `feedback' mechanism that shapes galaxy evolution (\citealt{Silk_and_Rees_1998}; \citealt{King2003}; \citealt{Fabian2012}; \citealt{Morganti2017}; \citealt{Harrison2018}; \citealt{Mosallanezhad2022}).

The line driving mechanism can significantly enhance the momentum transfer from radiation to matter, well beyond that of pure electron scattering. It was first proposed by \cite{Lucy1970} and subsequently developed by Castor, Abbott, \& Klein 
(\citeyear{Castor1975}, hereafter \citetalias{Castor1975}) to describe stellar winds from OB stars ($10^{4} \, {\rm K} < T_{\rm _{eff, OB}} < 5 \times 10^{4} \, {\rm K}$). This framework has been successfully applied to a variety of accretion disc systems, including AWDs and AGNs, where disc temperatures are comparable to those found in OB stars. In XRBs, line driving is less likely to play a dominant role since material near the compact objects tends to be very highly ionized.  In these systems, thermal and/or magnetic driving may be responsible for the observed outflows (\citealt{Miller2016}; \citealt{Higginbottom2017}; \citealt{Tomaru2020})\footnote{Recent observations of optical (e.g. \citealt{MataSanchez2018}; \citealt{MunozDarias2019}) and UV (e.g. \citealt{CastroSegura2022}) wind signatures in XRBs reveal lower-ionization material, implying that line driving may contribute to mass ejection in some of these systems.}.

In principle, radiation-driven winds will be generated in the vicinity of luminous sources whenever the outward force imparted via photon scattering and/or absorption exceeds the inward gravitational pull; the well-known Eddington limit \citep{Eddington1916} corresponds to the simplest scenario in which the radiative force arises purely via electron scattering, with $ L_{\rm Edd} $ representing the threshold luminosity above which radiation pressure overcomes gravity.
Extending this concept, line driving takes place in media that are not fully ionized, allowing photons to interact with bound electrons. These additional interactions can greatly amplify the overall radiative thrust. The factor by which the radiative force surpasses its value in an electron-only scattering regime is termed the force multiplier, $ \mathcal{M} $, a designation first introduced by \citetalias{Castor1975}. Under optimal ionization conditions, this multiplier can be as large as $ \mathcal{M} \simeq \mathrm{a~few} \times 10^3$ \citep{Gayley1995, Proga1998, Higginbottom2024}.

Numerical simulations of line-driven disc winds have been conducted both in systems hosting AWDs (\citealt{Proga1998}; \citealt{Pereyra1997, Pereyra2000}; \citealp{Dyda2018a, Dyda2018b}) and in the context of AGN-scale systems (\citealp{Proga2000, Proga2004, Liu2013, Nomura2016, Nomura2017, Mosallanezhad2019, Nomura2020, Dyda2024,Dyda2025}). A key element in such simulations is determining how the force multiplier is distributed throughout the outflowing material. More precisely, the force multiplier directly depends on the ionization state, $ \xi $, of the wind, which, in turn, is governed by radiative processes. At the same time, the radiation field itself is altered by its interactions with the outflowing gas. Consequently, matter and radiation are intricately coupled, each influencing the other in non-linear, complex ways within these line-driven outflows. 

Previous simulations have significantly simplified this complex interplay by employing the standard \citetalias{Castor1975} $k-\alpha$ parameterization, $\mathcal{M} = k\,t^{-\alpha}$. In this formulation, $t$ is the so-called optical depth parameter, which depends on local velocity gradients, while the exponent $\alpha$ captures the relative importance of optically thin versus optically thick lines. Finally, $k$ acts as a normalization factor that represents the overall effectiveness of line driving. In their AWD simulations, \cite{Proga1998} adopted $k=0.2$ and $\alpha=0.6$, guided by typical wind parameters in OB stars \citep{Gayley1995}. For their AGN models, \cite{Proga2000} likewise used $\alpha=0.6$, but replaced the constant $k$ with a function, $k(\xi)$, derived from the work of \cite{Stevens1990}. They also typically set the maximum allowed force multiplier to $ \mathcal{M}_{\rm max} \approx 4400$.

Despite early simulations successfully producing line-driven winds with mass-loss rates of $ \dot{M}_{\rm wind} \sim 10^{-4} \dot{M}_{\rm acc}$ \cite{Proga1998,proga1999}, certain discrepancies emerged \citep{Drew2000, Proga2002}. For instance, reproducing the strengths of wind-formed spectral lines in AWD systems required accretion rates that were higher than expected by a factor of 2--3. This hinted that some underlying assumptions may not be entirely accurate, although \citet{Proga03} found that theory and observations might be consistent after all. Later, more detailed radiative transfer calculations \citep{Sim2010, Higginbottom2014} revealed that multi-dimensional scattering effects in AGN outflows caused the winds to become overionized, disrupting effective line driving. These findings highlight a key concern: the force multipliers and ionization states assumed in these simulations -- based on the near-optimal conditions found in OB stars -- may not accurately reflect real astrophysical environments. 

To overcome these challenges, we \citep{Higginbottom2024} presented the first Monte Carlo radiation-hydrodynamic (MC-RHD) simulations, utilizing the \textsc{Pluto} hydrodynamics code coupled with the \textsc{Sirocco} radiative transfer code. These simulations incorporated detailed, multi-dimensional ionization state calculations via fully frequency-dependent radiative transfer, while self-consistently treating radiation transport and force multipliers in a unified framework. We focused on AWD systems because of their limited spatial dynamic range and simpler spectral energy distribution (SED) made them more manageable compared to AGN. In these simulations, we assumed an isothermal outflow at a constant temperature of $ T = 40{,}000\,\mathrm{K} $.
Our main finding was that the physical conditions in AWD environments were far less conducive to efficient line-driving than previously assumed. Unlike the conditions found in hot, single-star winds, the ionization states in these systems turned out to be significantly higher. 
Since these sources are only marginally luminous enough to sustain strong line-driven flows, this overestimation had profound implications. We found that the mass-loss rates were lower by approximately two orders of magnitude compared to previous estimates. Similarly, the synthetic spectra produced in our simulations did not match observational data. These results collectively raised serious questions about the viability of line-driving as the primary mechanism responsible for the winds observed in AWDs.

In our previous simulations, we assumed that the outflow was isothermal, a significant
simplification that ignores the temperature-dependent  aspects of ionization balance and radiative cooling/heating processes that govern the efficiency of line-driven winds.  For modelling winds in AGN, which is our ultimate goal and where the SEDs are inherently more complex and harder to parameterize than those of AWDs, the isothermal approximation becomes particularly problematic. To address this, we replace the isothermal assumption with  an ideal gas equation of state and directly solve the full set of energy equations. This enables us to resolve the spatial temperature structure of the outflow, driven by radiative and hydrodynamic heating/cooling processes. Critically, we also track the evolving SED within the outflow as it is locally attenuated, a feature essential for AGN applications where over-ionization can suppress line-driving. Our approach integrates these advancements with an enhanced 'macro-atom' method for radiative transfer \citep{lucy2002, lucy2003}, which more rigorously enforces energy conservation in the co-moving frame and results in a more accurate treatment of the reprocessing effect of the wind.

The remainder of this paper is organized as follows. In Section \ref{sec:method}, we begin with a brief overview of our previous simulations and then detail the newly developed methods used to perform 2.5D, multi-frequency RHD simulations with an ideal gas equation of state. This approach includes a full treatment of ionization, line-driving forces, and radiative heating/cooling rates. In Section \ref{sec:results}, we present the results of our simulations, some of which are directly compared against those from our previous isothermal models. Finally, in Section \ref{sec:Discussion}, we discuss our findings, examine the limitations of our calculations, and summarize our conclusions.

\section{Methods} \label{sec:method}

\subsection{Hydrodynamics incorporating radiative heating/cooling, and acceleration}
\label{sec:hydro}

Our MC-RHD simulations employ the publicly available Godunov-type hydrodynamics code \textsc{Pluto} (v4.4; \citealt{Mignone2007}), coupled with the Monte Carlo radiative transfer code \textsc{Sirocco} (\citealt{Long_and_Knigge2002}; extended by \citealt{Sim2005},  \citealt{Higginbottom2013}, \citealt{Matthews2015}, \citealt{Matthews2025}). The two codes are integrated via an operator-splitting formalism, in which  \textsc{Sirocco} supplies heating and cooling rates, as well as radiative accelerations. Our methodology is based on that described by \citealt{Higginbottom2024}, with the key difference being the explicit inclusion of the ideal gas equation of state and the solution of the full energy equation. 

Allowing for radiative forces, the hydrodynamic continuity and momentum equations can be written as:
\begin{equation} \label{eq:continuity}
\frac{\partial \rho}{\partial t} + \nabla \cdot (\rho \mathbf{v}) = 0,
\end{equation}
\begin{equation} \label{eq:momentum}
\frac{\partial (\rho \mathbf{v})}{\partial t} + \nabla \cdot (\rho \mathbf{v} \mathbf{v} + p \mathbf{I}) = -\rho \nabla \Phi + \rho \mathbf{g}_{\text{rad}}. 
\end{equation}
Here $ \rho $ is the gas density, $ \mathbf{v} $ is the velocity vector, $ p $ is the gas pressure, $ \mathbf{I} $ is the identity matrix, $ \Phi $ is the gravitational potential, and $ \mathbf{g}_{\rm rad} $ represents the radiative acceleration. 

In our previous simulations, we assumed the outflow to be isothermal, so these equations were supplemented with
\begin{equation} 
\label{eq:momentum_iso}
p = \rho c_{\rm iso}^{2}, 
\end{equation}
i.e. the isothermal equation of state. We adopted a fixed sound speed $ c_{\rm iso} = 24 \, \rm{km} \, \rm{s}^{-1} $, corresponding to a fixed temperature of $ T = 40,000 \, \text{K} $. With these assumptions, there was no need to solve the full energy equation.

In our new simulations, we relax the isothermal assumption and instead adopt the ideal gas equation of state
\begin{equation} \label{eq:momentum_iso}
p = n k_B T, 
\end{equation}
where $ n $ is the total particle number density, and $ k_B $ is the Boltzmann constant. We then do need to solve the full energy equation, which can be written as
\begin{equation} \label{eq:energy} 
\frac{\partial E}{\partial t} + \nabla \cdot \left[ (E + p) \mathbf{v} \right] = -\rho \mathbf{v} \cdot \nabla \Phi + \rho \mathbf{v} \cdot \mathbf{g}_{\text{rad}} + \rho \mathcal{L}.
\end{equation}
Here, $ E = \frac{1}{2} \rho |\mathbf{v}|^{2} + \rho e $ represents the total gas energy density, where $ e $ is the internal energy per unit mass, and $ \mathcal{L} $ is the net radiative heating/cooling rate per unit mass. We neglect relativistic effects and adopt a Newtonian gravitational potential, $ \Phi = - GM_{\rm _{WD}} / r $, where $ M_{\rm _{WD}} $ is the mass of the white dwarf, and $ G $ is the gravitational constant. For a monatomic ideal gas, the internal energy per unit mass, $ e $, is related to the pressure and density by $ e = p / [\rho (\gamma - 1)] $, where $ \gamma = 5/3$ is the adiabatic index.

In our simulations, \textsc{Pluto} is responsible for solving and advancing this coupled system of equations, with \textsc{Sirocco} providing updates for two critical terms: the radiative acceleration $ \mathbf{g}_{\text{rad}} $ and the net radiative heating/cooling rate $ \mathcal{L} $. Because the radiative transfer is much more expensive computationally than the hydrodynamical portion of the calculation, we carry out radiative transfer and ionization calculations at intervals significantly longer than the hydrodynamic time-step, specifically at $ \Delta t_{\rm _{RAD}} \gg \Delta t_{\rm _{HD}} $.  Typically, we set $\Delta t_{\rm _{RAD}} = 2 \, \rm{s} $, which corresponds to about $ 10^{3}\, \Delta t_{\rm _{HD}} $ in our simulations. For this to be a reasonable approach, the ionization state and radiation field within each cell remain approximately constant over the timescale $ t_{\rm _{RAD}} $. We have verified that our results are not sensitive to the exact value of $ \Delta t_{\rm _{RAD}} $ within this regime \footnote{We performed a simulation test using $ \Delta t_{\rm _{RAD}} = 0.2 \ \mathrm{s} $ and find that the wind mass-loss rate deviates by less than 5\% from $ \Delta t_{\rm _{RAD}} = 2.0 \ \mathrm{s} $ model, while the ionization structure shows only marginal variations at high latitudes. These results confirm that adopting $ \Delta t_{\rm _{RAD}} = 1000\,\Delta t_{\rm _{HD}} $ captures the essential dynamics without significant loss of accuracy--consistent with \cite{Higginbottom2024} -- and demonstrate that our conclusions are insensitive to the precise choice of $ \Delta t_{\rm _{RAD}} $ within this regime.}. Next, we cycle between hydrodynamic and radiative updates until the entire wind structure evolves to a quasi-steady state.
\footnote{By ``quasi-steady'', we mean a flow in which the mass-loss rate and overall geometry are no longer changing systematically with time. As previously found in RHD simulations of line-driven winds, the smaller-scale outflow structure is inherently time variable, even once this quasi-steady state is reached \citep{Proga1998, proga1999, Higginbottom2024}.}

In order to avoid abrupt temperature changes every $\Delta t_{\rm _{RAD}}$ (when exact new radiative heating/cooling rates become available), we apply approximate updates to $\mathcal{L}$ at each hydrodynamic time-step. These approximate updates are based on the estimated impact of the changing temperature and density on the radiative heating/cooling rates. We also apply a damping factor when we update $\mathcal{L}$ after each call to \textsc{Sirocco}. The way in which we calculate and update radiative heating and cooling rates is described in more detail in Section~\ref{subsection:hc_rates}.

Similarly, the radiative acceleration $\mathbf{g}_{\text{rad}}$ is updated approximately at each hydrodynamic time-step. Here, the approximate updates account for changes in the velocity field, while assuming that the ionization state of the flow stays approximately constant over a time-scale $\Delta t_{\rm _{RAD}} $. The way in which we calculate and update $\mathbf{g}_{\text{rad}}$ is described in more detail in Section~\ref{sec:grad}.

\subsection{Ionization and radiative transfer} 
\label{subsec:IRT}

After each simulation time interval of $ \Delta t_{_{\rm RAD}} $, \textsc{Sirocco} is activated and loads the latest snapshots of the density, velocity fields, and temperature calculated by \textsc{Pluto}, as well as the wind's ionization state from its previous run. Photon packets are then introduced into the outflow by both the accretion disc and the wind. These packets are tracked as they journey through the wind, where they may undergo attenuation due to bound-free and free-free opacity, electron scattering, and bound-bound interactions. The photon packets passing through each cell are used to build estimators that represent the radiation field within that cell. This information is then used to update the wind's ionization state and to calculate the radiative accelerations and radiative cooling/heating rates.

We begin by assuming that the disc initially has an effective temperature distribution following the standard \cite{Shakura_Sunyaev_1973} model

\begin{equation} \label{eq:T_vis}
    T_{\text{d, visc}}(R) = T_* \left( \frac{R_{_{\rm WD}}}{R} \right)^{3/4} \left( 1 - \frac{R_{_{\rm WD}}}{R} \right)^{1/4},
\end{equation}
where $ R = r \sin \theta $ denotes the cylindrical radius, and
\begin{equation}
T_{*} = \left( \frac{3 G M_{_{\rm WD}} \dot{M}_{\text{acc}}}{8 \sigma \pi R_{_{\rm WD}}^3} \right)^{1/4}.
\end{equation}

Here, $\dot{M}_{\rm acc}$ represents the accretion rate through the disc and, $ \sigma$ is the Stefan-Boltzmann constant. For the purpose of generating photon packets, the disc is divided into concentric annuli, each initially considered to radiate as a blackbody with an effective temperature given by $T_{\text{d, eff}} = T_{\text{d, visc}}(R)$. This setup determines both the number and the frequency distribution of the photon packets produced by each annulus. To be more precise, the plasma within the simulation domain itself generates photons through free–free, free–bound, and bound–bound processes.

As photons traverse the wind, they are employed to update three sets of estimators related to the radiation field: (i) the angle-averaged mean intensity $ J_\nu $, (ii) the direction-dependent ultraviolet (UV) flux $ F_{\rm _{UV,i}} $, and (iii) the radiative cooling and heating terms. Here, the index $ i $ denotes one of $ N_{\hat{n}} = 36 $ directions chosen to effectively sample all possible vectors in the $ \phi = 0 $ plane. The UV band is defined to span frequencies from $ 7.4 \times 10^{14} \, \text{Hz} $ to $ 3 \times 10^{16} \, \text{Hz} $. The flux estimator $ \vec{F}_{\rm _{UV,i}}$ is utilized by \textsc{Pluto} to compute the radiative accelerations (see Section \ref{sec:grad}), while the mean intensity estimator $ J_\nu $ is used to determine the ionization state of the gas and the corresponding relationship between the force multiplier and the optical depth parameter (refer to Section \ref{subsec:multiplier}). Additionally, the radiative cooling and heating terms are passed to \textsc{Pluto} to calculate the total cooling rate (see Section \ref{subsection:hc_rates} for more details).

After the photon transport phase is complete, \textsc{Sirocco} calculates the ionization state of the gas based on the local density, temperature, and the frequency-dependent intensity $ J_\nu $. The estimator for $ J_\nu $ inherently accounts for any attenuation of photon packets within the wind, as well as re-emission from the wind itself. Critically, it also includes photons that arrive in a cell from other regions of the wind, including those introduced via scattering events. This scattered and/or reprocessed component of the radiation field makes it much harder for material close to the center (e.g. a failed wind) to effectively shield material located further out.

Whenever \textsc{Sirocco} is called, we execute at least two iterations encompassing the entire process of photon generation, radiation transport, ionization equilibrium calculations and evaluation of the radiative cooling and heating terms. This iterative method enables us to account for the effects of irradiation on the temperature distribution of the disc. Some photon packets emitted by the central source and some that have been reprocessed in the outflow will unavoidably impact the disc's surface. In \textsc{Sirocco}, these photons can be managed in one of three ways: (i) by discarding them; (ii) by assuming they are absorbed and reprocessed; or (iii) by treating them as specularly reflected. In the simulations presented here, we choose option (ii). Consequently, after the first iteration, we update the effective temperature distribution across the disc such that
\begin{equation}
    \sigma T_{\text{d, eff}}^4(R) = \sigma T_{\text{d,visc}}^4(R) + F_{\text{irr}}(R),
\end{equation}
where $ F_{\text{irr}}(R) $ is the irradiating flux incident on the disc at radius $ R $.

\subsection{Radiative acceleration}
\label{sec:grad}

The net radiative acceleration in each cell, \( \mathbf{g}_{\text{rad}} \), is computed as a vector sum over \( N_{\hat{n}} \) directions uniformly distributed in the \( \phi = 0 \) plane, ensuring equal angular spacing between each direction,

\begin{equation}
\mathbf{g}_{\text{rad}} = \sum_{i=1}^{N_{\hat{n}}} g_{i} = \sum_{i=1}^{N_{\hat{n}}} \left[ 1 + \mathcal{M} (t_{i}) \right] \sigma_{\rm e} \frac{\vec{F}_{\rm _{UV,i} }}{c}.
\label{eq:grad}
\end{equation}
Here, $ \sigma_{\rm e} $ is the Thompson cross-section per unit mass, $ c $ is the light speed, and $ g_i $ and $ \vec{F}_{\rm _{UV,i}} $ represent the radiative acceleration and UV flux in direction $ i $, respectively. We conducted our simulations with a directional resolution of $ N_{\hat{n}} = 36 $. Tests at a higher resolution ($ N_{\hat{n}} = 72 $) showed no significant differences in the final results; the character of the wind remained nearly identical, with mass-loss rates differing by less than 1\% between the two simulations. 

In Equation~\ref{eq:grad}, $ \mathcal{M} (t_i) $ represents the force multiplier, which parameterizes the effective number of lines available to enhance the scattering coefficient. This quantity varies based on the ionization level of the material within each cell (see Section~\ref{subsec:multiplier}). Notably, the force multiplier depends on the optical depth parameter $ t_i $, defined as,
\begin{equation}
t_i = \sigma_{\rm e} \rho v_{\rm th} \left| \frac{d(\vec{v} \cdot \hat{n}_i)}{ds_i} \right|^{-1},    
\end{equation}
where $v_{\rm th} = \sqrt{2 k_{\rm _B } T / m_{\rm p }}$ is the thermal velocity of the gas, $ds_i$ represents an infinitesimal step along the direction $\hat{n}_i$, and $\left| d(\vec{v} \cdot \hat{n}_i) / ds_i \right|$ is the gradient of the velocity component projected in this direction. Therefore, in this study, by adopting an ideal gas equation of state, $ \mathcal{M}(t_i) $ becomes direction-dependent and is sensitive to the instantaneous local velocity field as well as the temperature distribution.
It is worth highlighting that even though the ionization state of the flow is updated only every $\Delta t_{\rm _{RAD}}$, the optical depth parameter -- and hence the force multiplier -- are adjusted continuously (at each hydrodynamic time step) to account for the changing velocity field in each cell.

These conceptually simple expressions capture the complex nonlinear coupling between matter and radiation in line-driven flows. The radiation field and ionization state depend on the outflow dynamics -- which in turn control the density, velocity fields, and temperature. Conversely, the dynamics are influenced by the radiation field and ionization state, which affect $ {F}_{\rm _{UV,i}} $, $ \mathcal{M}(t_i) $, and the net radiative rate.

\subsection{Generating a lookup table for the force multiplier} \label{subsec:multiplier}

We follow \cite{Higginbottom2024} in utilizing the approach described by \cite{Parkin_Sim_2013} to generate updated look-up tables for the local force multiplier, considering the SED and ionization parameter within each cell. We do not employ \textsc{Sirocco} directly for this task because calculating precise force multipliers requires a more extensive line list than the one \textsc{Sirocco} uses. \textsc{Sirocco}'s line list is sufficient for obtaining the accurate estimate of $ J_\nu $ needed to determine the ionization state, but not for the detailed force multiplier calculations. Essentially, we provide the ionization state and the local estimate of $J_\nu$ to an independent code that has access to over 450,000 spectral lines (see \citealt{Parkin_Sim_2013} for further details). For each transition, the code begins by calculating the quantity
\begin{equation}
    \eta_{u,l} = \frac{hc}{4 \pi} \frac{n_{l} B_{l,u} - n_{u} B_{u, l}}{\sigma_e \rho v_{\text{th}}},
\end{equation}
where $ u $ and $ l $ refer to the upper and lower level of the relevant transition, respectively. Here, $ n_u $ and $ n_l $ are the upper and lower level number densities of ions supplied by \textsc{Sirocco} and $ B_{l,u} $ and $ B_{u,l} $ are the usual Einstein coefficients for absorption and stimulated emission, respectively. The force multiplier $ \mathcal{M}(t) $ is determined as a weighted sum of contributions from all line transitions
\begin{equation}
    \mathcal{M}(t) = \sum_{\text{lines}} \Delta \nu_{\rm _D} \frac{ J_\nu}{J} \frac{1 - \exp(-\eta_{u,l}\, t)}{t},
\end{equation}
where $ \Delta \nu_{\rm _D} = \nu_{_0} v_{\rm th} / c $ represents the the Doppler width of each line with $ \nu_{_0} $ as the central frequency of the line. 

\subsection{Heating and cooling rates} \label{subsection:hc_rates}

After each call to \textsc{Sirocco}, i.e. every $ \Delta t_{\rm _{RAD}} $, we pass updated radiative heating and cooling rates back to \textsc{Pluto}. However, as already noted in Section~\ref{sec:hydro}, we do not take the radiative heating and cooling rates to be constant between these calls. Instead, we apply approximate updates to these rates at each hydrodynamic time step in order to allow for rapid changes in the local density, temperature and ionization state. These approximate updates are essentially a way to interpolate the radiative heating and cooling rates across $\Delta t_{\rm _{RAD}}$. The purpose of this technique is to facilitate convergence by preventing unphysically abrupt changes in the simulation after calls to \textsc{Sirocco}.

In practice, we calculate approximate updates to $\rho \mathcal{L} $ at each hydrodynamic time step by using slightly modified versions of the analytic heating and cooling rate equations given by \citealt{Higginbottom2018} (which are themselves based on \citealt{Blondin1994}; also see \citealt{Higginbottom_Proga_2015, Higginbottom2017}). In these equations, the net radiative heating and cooling rate is broken up into five heating and cooling terms: Compton heating ($ H_{\rm c} $) and Compton cooling ($ \Lambda_{\rm c} $), X-ray (photoionization + Auger) heating rate ($ H_{\rm x} $), bremsstrahlung cooling ($ \Lambda_{\rm b}  $), and cooling via line emission ($ \Lambda_{\rm l} $). All five heating and cooling rates are then combined to give an estimate of the total net rate
\begin{equation}
    \rho \mathcal{L} = n_{\rm _H} \left( n_{\rm e} H_{\rm c} + n_{\rm _H} H_{\rm x} - n_{\rm e} \Lambda_{\rm c} - n_{\rm e} \Lambda_{\rm b} - n_{\rm e} \Lambda_{\rm l} \right),
\end{equation}
where $ n_{\rm _H} $ and $ n_{\rm e} $ are the number densities of hydrogen and electrons, respectively. Following \cite{Higginbottom2018}, we approximate the electron-to-hydrogen density ratio as
\begin{equation}
   \frac{n_{\rm e}}{n_{\rm _H}} = \begin{cases}
    2.5 \times 10^{-52} \, T^{12.3}   & \, T < 1.5 \times 10^{4}\, \text{K} \\
    2.5 \times 10^{-3.8} \, T^{0.86} & \, 1.5 \times 10^{4} \, \text{K} \leq T < 3.3 \times 10^{4}\, \text{K} \\
    1.21 & \,  T \geq 3.3 \times 10^{4} \, \text{K}.
\end{cases} 
\end{equation}

The heating and cooling terms can then be expressed as,
\begin{equation} \label{heat_comp}
    H_{\rm c} = K_{H_{\rm c}} \left[ 8.9 \times 10^{-36} \, \xi \, T_{\rm x} \right] \, ({\rm erg}\, {\rm s}^{-1}\, {\rm cm}^{3} ), 
\end{equation}
\begin{equation} \label{cool_comp}
    \Lambda_{\rm c} = K_{\Lambda_{\rm c}} \left[ 8.9 \times 10^{-36} \, \xi \, (4T) \right] \, ({\rm erg}\, {\rm s}^{-1}\, {\rm cm}^{3} ),
\end{equation}
\begin{equation} \label{heat_xray}
    H_{\rm x} = K_{H_{\rm x}} \left[  1.5 \times 10^{-21} \, \xi^{1/4} T^{-1/2}  \right] \, ({\rm erg}\, {\rm s}^{-1}\, {\rm cm}^{3} ),
\end{equation}
\begin{equation} \label{cool_brem}
    \Lambda_{\rm b} = K_{\Lambda_{\rm b}} \left[ 3.3 \times 10^{-27} \, T^{1/2} \right] \,  ({\rm erg}\, {\rm s}^{-1}\, {\rm cm}^{3})
\end{equation}

\begin{equation} \label{cool_line}
    \Lambda_{\rm l} = K_{\Lambda_{\rm l}}  \left[ 1.0 \times 10^{-16} \, \frac{\exp \left(-1.3 \times 10^{5} / T \right)}{T \sqrt{\xi}} + \mathcal{K}(T) \right] \, ({\rm erg}\, {\rm s}^{-1}\,\, {\rm cm}^{3}),
\end{equation}
where
\begin{equation}
    \mathcal{K}(T) = \begin{cases} \label{K_T}
      5.0 \times 10^{-27} T^{1/2} & T < 10^4 \, \text{K} \\
      1.0 \times 10^{-24} & 10^4 \, \text{K} < T < 10^7 \, \text{K} \\
      1.5 \times 10^{-17} T^{-1} & T > 10^7 \, \text{K}.
\end{cases}
\end{equation}
The quantity $\xi \left[ = 4\pi F_{\rm ion} n_H^{-1} \right]$ in these equations is the so-called ionization parmeter (where $F_{\rm ion}$ denotes the Hydrogen-ionizing flux).

These analytic expressions -- without the ``$K$ prefactors'' -- were originally designed to approximate the heating and cooling rates for material irradiated by a Bremsstrahlung SED. However, the scalings of the terms with temperature, density and ionization parameter should be fairly robust to SED changes, 
and it is only these scalings that our approximate updating method actually relies on. 

Specifically, the analytic expressions are implemented in \textsc{Pluto} and used to continually update the radiative heating and cooling rates at each hydrodynamic time step. However, they are recalibrated after each call to \textsc{Sirocco} by adjusting the $K$ prefactors so that the approximate rates match the {\em actual} rates (as calculated by \textsc{Sirocco}). 

To prevent abrupt changes in the heating and cooling rates, we apply a numerical damping factor during the recalibration process after each ionization and raditive transfer (IRT) step. While the simulation is expected -- and observed -- to converge to a quasi-steady state, the exact trajectory to this state is not critical. The numerical damping facilitates convergence without impacting the final results. We have tested this approach by using various damping factors and confirmed its robustness. For our final runs, we used a damping factor of 0.9, meaning the calibration factors can change by no more than 10 \% after each IRT step.

\subsection{System parameters and computational set-up} \label{subsec:initial}

The system parameters and numerical set-up we adopt largely follow those used for Model HK22D by \cite{Higginbottom2024}, which they used as their fiducial reference model. However, we summarize the key points here briefly for completeness. Our numerical grid is set up in spherical polar coordinates $ (r, \theta, \phi) $. The computational domain is defined in two dimensions, spanning $ r_{\rm min} \leq r \leq r_{\rm max} $ and $ 0 \leq \theta \leq \pi/2 $, covering one quadrant of the environment near the white dwarf for hydrodynamical calculations. We set $ r_{\rm min} = r_{\rm _{WD}} $ and $ r_{\rm max} = 10 \, r_{\rm _{WD}} $, where $ r_{\rm _{WD}} $ is the radius of the white dwarf \footnote{Our primary goal here is to compare our ideal equation‐of‐state and macro‐atom mode simulations with the isothermal wind models of \cite{Higginbottom2024}, which were computed over a radial domain of $ 10 \, r_{\rm _{WD}} $. To assess the impact of domain size, we performed a test run extending the outer boundary to $ r_{\rm max} = 50 \, r_{\rm _{WD}} $ and adopting the midplane density of $ \rho(r,\theta=\pi/2) \propto r^{-\alpha} $ with $ \alpha=0.5 $. The simulation reaches a steady state after approximately $ 400 \, \mathrm{s} $, and the average mass‑loss rate matches that of our fiducial model closely, confirming that extending the radial domain does not significantly alter the global wind properties.}. The parameters are fixed as $ M_{\rm _{WD}} = 0.6 \, M_{\odot} $ and $ r_{\rm _{WD}} = 8.7 \times 10^{8} \, {\rm cm} $.

The $ r $-$ \theta $ plane is divided into zones as follows: in the $ r $-direction, the domain is discretized into 128 zones with a geometric progression such that $ \mathrm{d} r_{i} / \mathrm{d} r_{i+1} = 1.05 $. Similarly, in the $ \theta $-direction, the domain consists of 96 zones with $ \mathrm{d} \theta_{j} / \mathrm{d} \theta_{j+1} = 0.95 $. This spacing ensures the highest resolution in the wind-launching region, particularly near the accretor and the disc plane. The only ``net'' radiation source in our model is the accretion disc: there is no central source, and the wind only reprocesses disc photons. As discussed and shown by \cite{Higginbottom2024}, this is the most optimistic assumption for successful line-driving. The accretion rate is set to $ \dot{M}_{\text{acc}} = \pi \times 10^{-8} \, M_{\odot} \, \text{yr}^{-1} $, which is at the high end of the typical range for wind-driven accretion-disc systems, such as nova-like variables (\citealt{Howell_Mason_2018}). 

\begin{table*}
    \centering
    \renewcommand{\arraystretch}{1.3} 
    \setlength{\tabcolsep}{8pt}      
    \caption{Parameters adopted in the simulations and some derived quantities. For each simulation, we provide the model designation (column ‘Model Name’), its Sirocco mode, the equation of state, the equivalent models in HK, the accretion rate (‘$ \dot{M}_{\rm acc} $’), the wind mass-loss rate (‘$ \dot{M}_{\rm wind} $’), and the velocity of the fast parts of the wind (‘$ v_{r} $’). Full parameter files are available in the code repository (see {\sl Data Availability}).
    } 
    \label{tab:models}
    \begin{tabular}{|c|c|c|c|c|c|c|c|}
        \hline
        \textbf{Name} & \textbf{SIROCCO Mode} & \textbf{Equation of State} & \textbf{Wind Temperature} & \textbf{Eqv. Model}  & \boldmath\(\dot{M}_{\text{acc}}\) & \boldmath\(\dot{M}_{\text{wind}}\) & \boldmath\(v_r\) \\ 
        & & & $ \left( \mathrm{K}\right) $ & \textbf{in HK24} & $ \left( M_{\odot} \text{ yr}^{-1} \right) $ & $ \left( M_{\odot} \text{ yr}^{-1} \right) $ & $ \left( \text{km s}^{-1} \right) $ \\ 
        1 & 2 & 3 & 4 & 5 & 6 & 7 & 8 \\ \hline \hline
        \textbf{Model A} & \textbf{Hybrid Macro-Atom} & \textbf{Ideal} & \textbf{Variable} & \textbf{...} & \boldmath\(\pi \times 10^{-8}\) & \(\mathbf{6.3 \times 10^{-14}}\) & \textbf{1900} \\ 
        Model B & Hybrid Macro-Atom & Isothermal & 40,000 & ... & \(\pi \times 10^{-8}\) & \(5.6 \times 10^{-14}\) & \(1250\) \\ 
        Model C & Classic & Ideal & Variable & ... & \(\pi \times 10^{-8}\) & \(3.6 \times 10^{-14}\) & \(1700\) \\ 
        Model D & Classic & Isothermal & 40,000 & HK22D & \(\pi \times 10^{-8}\) & \(4.6 \times 10^{-14}\) & \(1700\) \\ \hline \hline
    \end{tabular}
    \vspace{1em} 
    \begin{flushleft}
        \textbf{Note:} Here, HK24 refers to our previous simulations, i.e., \citealt{Higginbottom2024}.  
    \end{flushleft}
\end{table*}

We assume a disc that is geometrically thin (\citealt{Shakura_Sunyaev_1973}; SSD), meaning all the radiation from the disc originates from the system's mid-plane at $ \theta = \pi/2 $. Since the disc serves as the mass reservoir for the outflow, we set the density at mid-plane to a fixed value and maintain this constant throughout the simulations. Ideally, this density, $ \rho_\mathrm{d} $, should correspond to the upper regions of the disc's atmosphere, just below where the wind is launched. For our simulations, $ \rho_\mathrm{d} $ must be set sufficiently high to ensure that the critical and sonic points of the outflow fall within the simulation domain, but not so high that the hydrostatic regions within the domain become optically thick. We typically choose $ \rho_\mathrm{d} = 10^{-9} \, \text{g cm}^{-3} $ to achieve this balance. We tested variations in \(\rho_{\rm d}\) by factors of two and according to the standard SSD model, finding no significant changes in the wind solution; the global wind properties remain robust. Therefore, as long as $ \rho_\mathrm{d} $ is chosen to correctly capture the sonic point, while avoiding excessive optical thickness at the base of the wind, our results remain relatively insensitive to its precise value. The initial density distribution across the grid is determined through hydrostatic equilibrium in the latitudinal direction, expressed as:
\begin{equation}
    \rho(r, \theta) = \rho_\mathrm{d} \exp \left( -\frac{G M_{\rm _{WD} }}{2c^2_{\text{s}} r \tan^2 \theta }  \right). 
\end{equation}
Here, $ c_{\text{s}} = \sqrt{\gamma k_{\rm _B} T / (\mu m_{\rm p})} $ is the sound speed, and $ \mu = 0.6 $, and $ m_{\rm p} $ are the mean molecular weight, and proton mass, respectively. The initial temperature distribution is set via Equation (\ref{eq:T_vis}), i.e., $ T(r , \theta) = T_{\text{d,visc}} $. The initial pressure is given by $ p = \rho c_{\rm s}^{2} / \gamma $. For the initial velocity, we assume $ v_r = v_{\theta} = 0 $, and $ v_{\phi} = v_{\rm k} $, where $ v_{\rm k} $ is the Keplerian velocity. For isothermal simulations, we fix the temperature to $ T = 40,000 \, \text{K} $, which corresponds to an isothermal sound speed of $ c_{\rm iso} = 24 \, \text{km} \, \text{s}^{-1} $.

In all of our simulations, we impose a density floor of $ \rho_{\rm floor} = 10^{-24} \, \text{g}\, \text{cm}^{-3} $. If the density in a cell falls below this threshold, it is reset to $\rho_{\rm floor}$ while conserving both momentum and energy. 

After each hydrodynamic time step, $ \Delta t_{\rm _{HD}}$, the mid-plane temperatures and densities are reset to $T_{\rm d, \, \text{visc}}$ and $ \rho_{\rm d} $ to $ 10^{-9} \, \text{g}\, \text{cm}^{-3} $, respectively. After resetting the mid-plane density, we also update the velocity in the $\theta$ direction $v_\theta$ to ensure momentum conservation in this direction. 

\begin{figure*}
    \centering
    \includegraphics[width=0.75\textwidth]{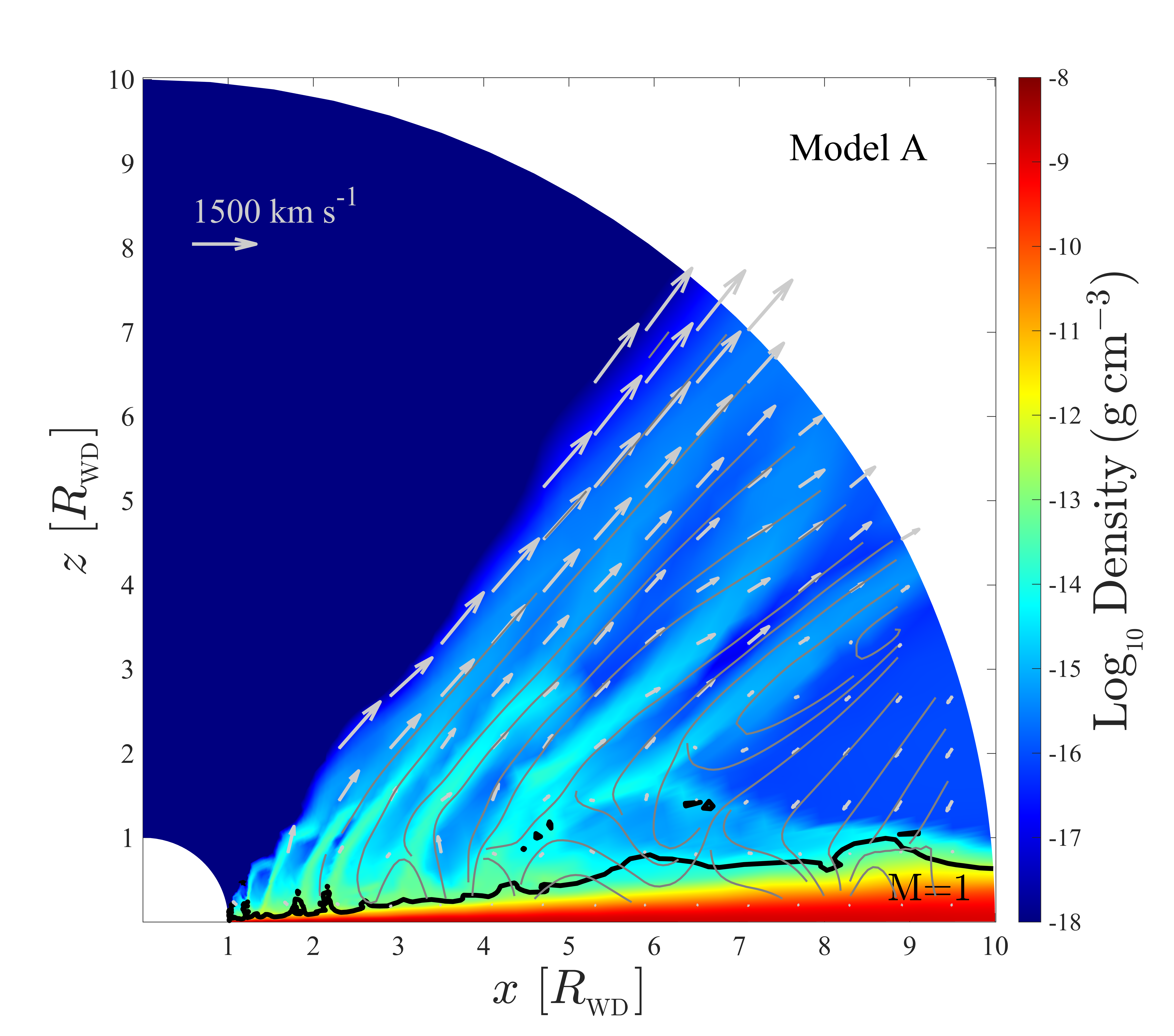}
    \caption{The density and poloidal velocity fields for the fiducial model at \( t = 850\,\mathrm{s} \) are shown. The colormap depicts the logarithmic density, while overlaid velocity vectors (normalized to a peak poloidal velocity of \( v_{p} = 1500\,\mathrm{km}\,\mathrm{s}^{-1} \)) illustrate the flow structure. Grey lines denote streamlines, and the solid black line marks the Mach 1 surface. For a version of the figure with logarithmically scaled \( x \) and \( z \) axes, see the top right panel of Fig.~\ref{fig:loglog_bars}. A line-driven disc wind is produced, with the bulk of the wind confined to a \( 25^\circ \) wedge spanning polar angles from \( 40^\circ \) to \( 65^\circ \).}
    \label{fig:model_A}
\end{figure*}

We adopt outflow boundary conditions at the innermost and outermost radial cells. The boundary at $ \theta = 0 $ is treated as axisymmetric, while reflection symmetry is assumed at the mid-plane, $\theta = \pi/2 $. For all of our simulations, we employ \textsc{Pluto}'s standard hydrodynamics module with linear reconstruction. Time integration is performed using a second-order Runge-Kutta (\textsc{RK2}) scheme, with a Courant-Friedrichs-Lewy (\textsc{CFL}) number of 0.4. We adopt the Harten–Lax–van Leer (\textsc{HLL}) approximate Riemann solver, which is robust in handling discontinuities.

\begin{figure*}
    \centering
    \includegraphics[width=\textwidth]{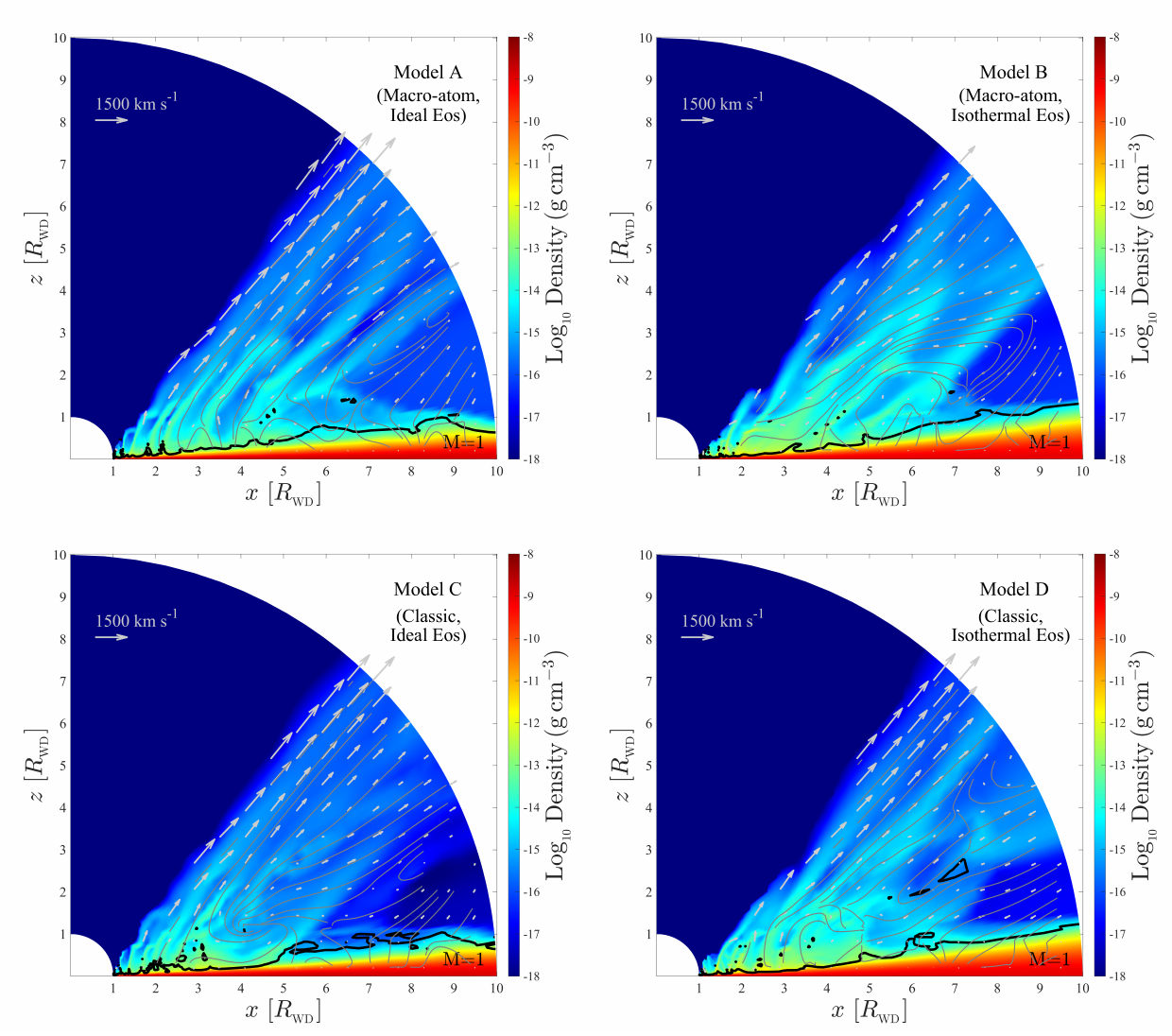}
    \caption{The density and poloidal velocity fields for Model A (top-left panel), Model B (top-right panel), Model C (bottom-left panel), and Model D (bottem-right panel; similar to Model HK22D of \citealt{Higginbottom2024}) at $ t = 850 \, \rm{s} $. The colourmap represent the logarithmic density, while overlaid velocity vectors (normalized to a peak poloidal velocity of \( v_{p} = 1500\,\mathrm{km}\,\mathrm{s}^{-1} \)) illustrate the flow structure. Grey lines denote streamlines, and the solid black line marks the Mach 1 surface. Animated versions of each panel can be found in the supplementary material.} 
    \label{fig:all_densities}
\end{figure*}

To initialize the radiative force, we use \textsc{Sirocco} to compute the ionization state, direction-dependent fluxes, and radiative cooling/heating terms following the procedure outlined in Sections~(\ref{sec:hydro})–(\ref{subsection:hc_rates}). This allows us to initialize the radiation force and radiative cooling/heating in \textsc{Pluto} for the first time step. Each time \textsc{Sirocco} is called -- i.e. after each $\Delta t_{\rm _{RAD}}$ -- we run two ``ionization cycles"  \cite[c.f.][]{Matthews2025}, each of which tracks $10^7$ energy packets through the computational domain. All simulations were evolved for a total of $1500$~s, at which point they had all reached quasi-steady states.

As discussed by \cite{Matthews2025}, \textsc{Sirocco} offers two modes of operation for radiative transfer: the Hybrid macro-atom mode, which is generally more physically accurate, and the Classic mode. In the latter, Monte Carlo packets are allowed to lose energy due to continuum opacity as they traverse the flow, with this energy taken up by packets emitted separately by the wind. In the former, near radiative equilibrium and co-moving frame energy conservation are instead enforced rigorously at the point of interaction. In addition, the hybrid macro-atom mode makes use of the macro-atom formalism \citep{lucy2002,lucy2003}, in our case treating both H and He as macro-atoms and the remaining metals as `simple atoms' within a two-level approximation for line transfer \citep[see][Section 3, for more details]{Matthews2025}. One beneficial aspect of this approach is that the bound-free continua of H and He ions, as well as their recombination cascades, are treated more accurately. In our previous radiation-hydrodynamic simulations, we only employed the Classic mode. We now run each of our isothermal and ideal gas simulations using both radiative transfer modes, which allows us to test if any of our science conclusions are sensitive to the details of the methods. Is also acts as a proof-of-concept, since this is to our knowledge the first time a radiation-hydrodynamics simulation of line-driven winds has used a version of the macro-atom formalism.

\begin{figure*}
    \centering
    \includegraphics[width=0.8\textwidth]{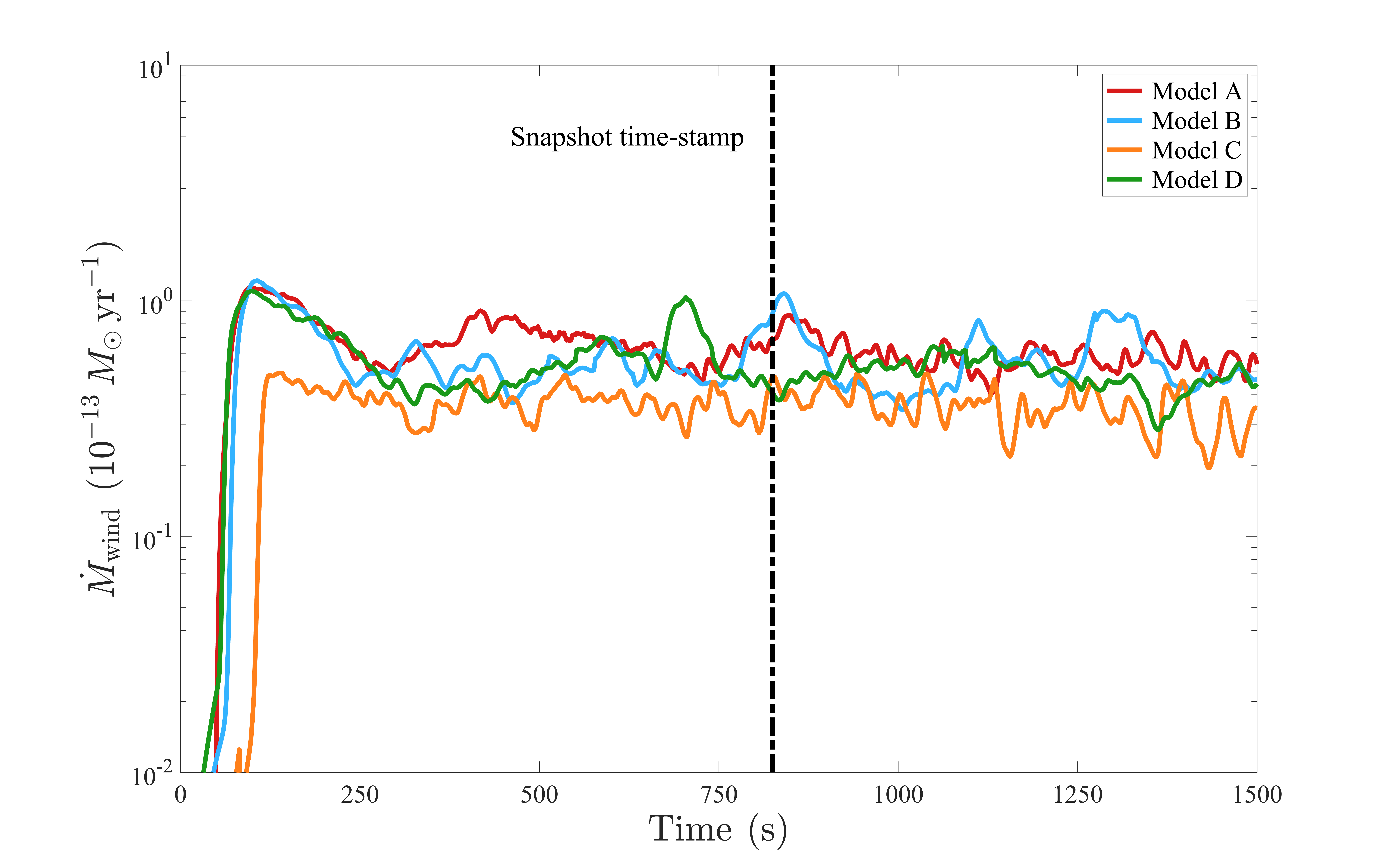}
    \caption{The time evolution of the wind mass-loss rate ($\dot{M}_{\mathrm{wind}}$) through the outer boundary for all four models. A vertical dash-dotted line marks the timestamp corresponding to the representative snapshot shown in subsequent figures. The high-density disc region is excluded in the calculation of $\dot{M}_{\mathrm{wind}}$}
    \label{fig:mass_outflow}
\end{figure*}

\section{Results} 
\label{sec:results}

We performed four simulations to investigate the impact of the ideal versus isothermal equations of state while employing the two different \textsc{Sirocco} modes described above: the Hybrid macro-atom and the Classic modes. The detailed parameters for these models are listed in columns 1--6 of Table \ref{tab:models}. Specifically, Models A and B utilize the \textsc{Sirocco} Hybrid macro-atom mode to directly compare the behavior of the ideal and isothermal cases, whereas Models C and D are executed using the Classic \textsc{Sirocco} mode. Model D in our simulations is identical in setup to Model HK22D from \cite{Higginbottom2024}. The key parameters describing our simulations are summarized in  Table~\ref{tab:models}. Model A represents our most physically complete simulation of a line-driven accretion disc wind to date.

Fig.~\ref{fig:model_A} displays a snapshot of the density distribution, overlaid with the poloidal velocity $ v_{p} = (v_{r}^2 + v_{\theta}^2)^{1/2} $, for our fiducial model. This plot corresponds to $ t = 850 \, \mathrm{s} $ of simulation time, approximately one-fifth of the sound-crossing time scale, at which point the simulation has reached a quasi-steady state. In this figure, the colormap represents the logarithmic density, while the overlaid velocity vectors, normalized to a peak poloidal velocity of $ v_{p} = 1500 \, \mathrm{km} \, \mathrm{s}^{-1} $, illustrate the wind structure. Grey lines denote streamlines, and the solid black line marks the Mach 1 surface. A line-driven disc wind is produced, with the majority of the wind confined to a $ 25^\circ $ wedge, spanning polar angles from $ 40^\circ $ to $ 65^\circ $. 

\begin{figure*}
    \centering
    \includegraphics[width=0.65\textwidth]{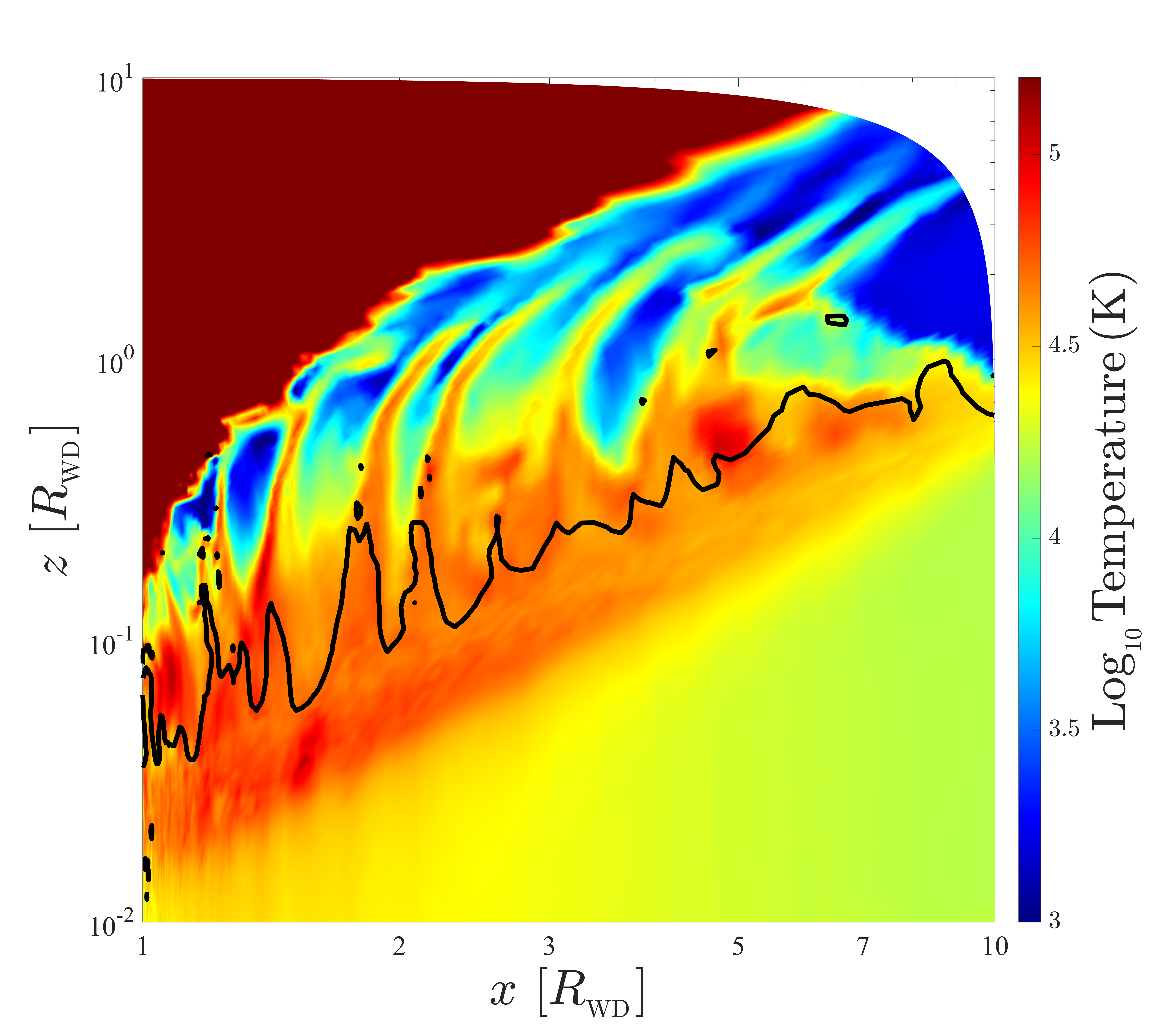}
    \caption{Two-dimensional temperature distribution from \textsc{Pluto} hydrodynamic simulation snapshot of our fiducial model, shown on a logarithmic color scale. The Mach~1 surface is indicated by the solid black line. Near this surface, temperatures reach a peak of $T \approx 5\times10^4\ \mathrm{K}$, in good agreement with the isothermal wind assumption of $T = 4\times10^4\ \mathrm{K}$ used in isothermal models.}

\label{fig:temperature}
\end{figure*}

Fig.~\ref{fig:all_densities} displays snapshots of the density distribution for all four models at a similar time ($t = 850 \, {\rm s}$, roughly one-fifth of the sound-crossing time scale). All simulations have reached a quasi-steady state by this point (see Fig.~\ref{fig:mass_outflow} and discussion below). In each panel, we also show the corresponding poloidal velocity field, $ v_{p} $. 

All four models produce similar line-driven disc winds, with comparable outflow geometries and velocity fields. In particular, the majority of the outflow is always confined to a $ 25^\circ $ wedge spanning polar angles from $ 40^\circ $ to $ 65^\circ $. The highly structured and variable nature of the outflow -- a feature already noted in the earliest simulations by \cite{Proga1998} -- is also consistent across all simulations. 

Our new benchmark model, Model A, exhibits a slightly higher wind velocity compared to the other cases (c.f. Table~\ref{tab:models}. Moreover, the inclusion of cooling and heating terms in the energy equation for the ideal runs (see the top and bottom left panels of Fig.~\ref{fig:all_densities}) results in a narrower high-density region near the equatorial plane. As a consequence, the lower-density environment in these models shifts the sonic (Mach~1) surface somewhat closer to the equator.

Perhaps the most important product of our simulations is the mass loss rate through the outer radial boundary (i.e., \( r_{\rm max} \)). The total mass outflow rate at this boundary is computed as
\begin{equation} \label{eq:mass_outflow}
    \dot{M}_{\rm wind} = 4 \pi r_{\rm max}^{2} \int_{0}^{85^{\circ}} \rho \, {\rm max} \left( v_{r}, 0 \right) \sin \theta \, d\theta.
\end{equation}
The upper limit of the integration here is set at $85^\circ$ to exclude the dense and essentially hydrostatic disc atmosphere. 

 Fig.~\ref{fig:mass_outflow} illustrates the time evolution of the wind mass-loss rate ($\dot{M}_{\mathrm{wind}}$) for all four models. In all simulations, the mass-loss rate stabilizes to be approximately steady within $200\,\text{s}$. The remaining fluctuations in the mass-loss rates beyond this point are associated with the density structures apparent in Fig.~\ref{fig:model_A}, which remain time variable even in the quasi-steady state. The vertical dash-dotted line in Fig.~\ref{fig:mass_outflow} marks the timestamp corresponding to the representative snapshot shown in subsequent figures.

Fig.~\ref{fig:mass_outflow} shows that the mass-loss rates predicted by all simulations are broadly consistent, to within roughly the scatter expected from their intrinsic variability. More quantitatively, Column~7 of Table~\ref{tab:models} lists the time-averaged $\dot{M}_{\mathrm{wind}}$ over the interval $300\,\text{s} \leq t \leq 1500\,\text{s}$.  All four simulations produce values within roughly a factor of two of each other. This similarity in mass-loss rates demonstrates that the global properties of the wind are not particularly sensitive to the treatment of thermodynamics (isothermal versus full thermal balance) or to the details of the radiative transfer mode (Classic versus Hybrid macro-atom).
  
Crucially, the mass-loss rate in Model~A remains a factor of 100 lower than the values found in previous simulations that did not treat the multi-dimensional radiative transfer and ionization in detail \cite[e.g.,][]{Pereyra1997, Pereyra2000, Proga1998, proga1999, Dyda2018a, Dyda2018b}. Thus the challenge to line-driving highlighted by our previous work cannot be ascribed to the isothermal approximation adopted there, and nor is it sensitive to the mode of radiative transfer used within \textsc{Sirocco}. We will return to this issue in Section~\ref{sec:Discussion}.

\begin{figure*}
    \centering
    \includegraphics[width=\textwidth]{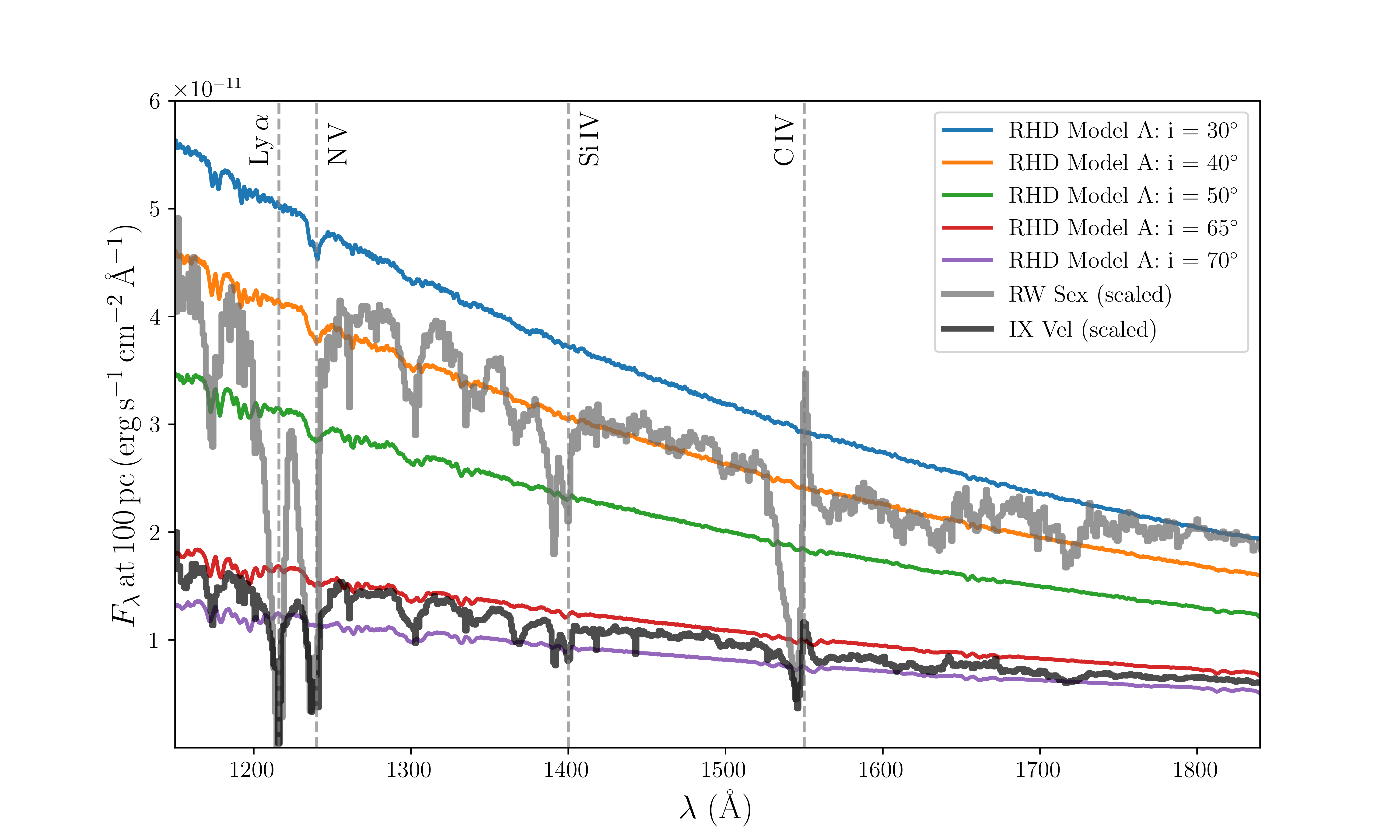}
    \caption{Synthetic UV spectra generated from a snapshot of the fiducial model for a range of inclination angles using \textsc{Sirocco}. We also show the ultraviolet spectra of two proto-typical high-state AWDs: RW Sex ($i \simeq 30^\circ$) and IX Vel ($i \simeq 65^\circ$). The spectra are normalized such that the flux levels correspond to a system observed at 100~pc. The positions of the Lyman limit and several key UV resonance lines are marked with vertical grey dashed lines.}

\label{fig:all_spectra}
\end{figure*}

\begin{figure*}
    \centering
    \includegraphics[width=\textwidth]{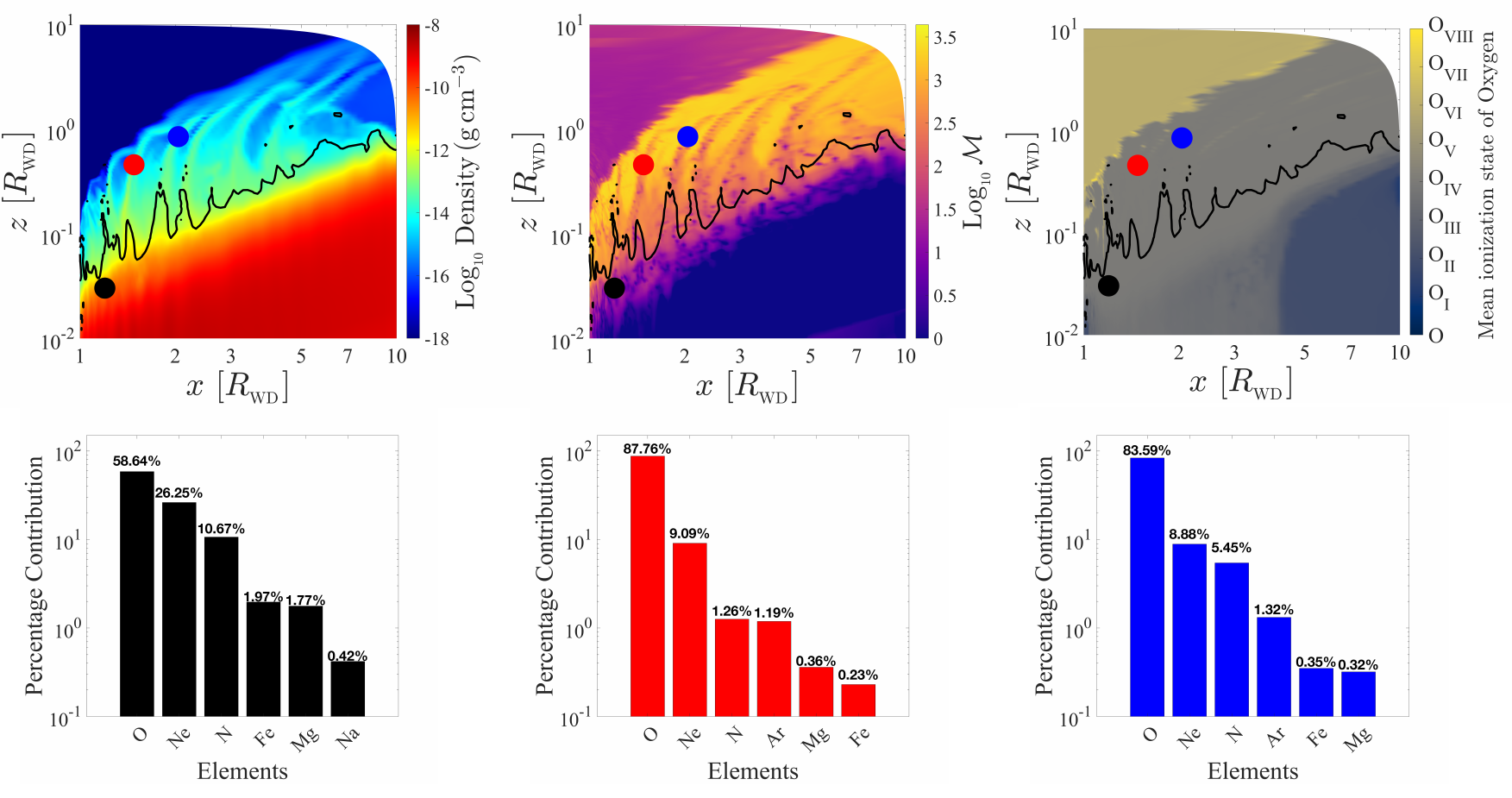}
    \caption{
    \textbf{Top panels:} Left panel displays the density map with logarithmic axes, middle panel presents the distribution of the force multiplier ($\mathcal{M}$), right panel illustrates the mean ionization state of oxygen in our fiducial model. The solid black line marks the location of the Much~1 surface. Black, red, and blue dots indicate the positions of three representative cells selected for analysis in the simulation domain. 
    \textbf{Bottom panels:} Percentage contributions of various elements to the total force multiplier at these representative points.
    }
\label{fig:loglog_bars}
\end{figure*}

\subsection{The wind temperature distribution} 
\label{subsec:temperature}

Fig.~\ref{fig:temperature} displays the temperature structure of the outflow for Model A. Near the Mach 1 surface, where the wind is launched, we find $T \simeq 5\times10^4$~K. This is very similar to the value adopted in our isothermal models, $T = 4\times10^4$~K, which accounts for the good agreement between the models.

It is worth noting that there are some regions in the simulation -- notably near the tips of the finger-like density structures -- where advective heat transport becomes dominant over radiative or adiabatic heating/cooling. Strictly speaking, the assumption of instantaneous ionization equilibrium we make in \textsc{Sirocco} also breaks down in these regions, since under these conditions the flow time-scale is likely to become shorter than the recombination time scale. The ionization state would then become ``frozen-in'' \citep[see e.g.][]{drew-verbunt}. However, this is unlikely to be a serious issue for the simulations here, since these regions are far from the wind-launching zone, and their temperature and ionization structure remains relatively constant (and physically reasonable) in any case. 

\subsection{Ionization state, force multipliers and driving species} 
\label{subsec:ionization_driving species}

As previously discussed by \cite{Higginbottom2024}, the relatively low mass-loss rates in our simulations are caused by the outflow becoming over-ionized. The actual outflows we observe in high-state AWDs do not seem to suffer from this problem.  This discrepancy can be appreciated by comparing observed and predicted spectra. Fig.~\ref{fig:all_spectra} shows the ultraviolet spectra of two proto-typical high-state AWDs (RW~Sex, $i \simeq 30^{\circ}$ and IX~Vel, $i \simeq 65^{\circ}$; HST program 14637 [PI: Long]). The dominant features here are strong, wind-formed resonance lines associated with N~{\sc v}~1240~\AA, Si~{\sc iv}~1400~\AA~ and C~{\sc iv}~1550~\AA.  These features are weak or absent in the spectra predicted by \textsc{Sirocco} for Model A for the same inclinations. This reflects the ionization state of the model wind, which favours higher ionization stages for each these species. 

In Fig.~\ref{fig:loglog_bars}, we provide a more detailed picture of the ionization state of the wind and its impact on the force multiplier. In the top row of Fig.~\ref{fig:loglog_bars}, we plot the distribution of density (left panel), force multiplier (middle panel) and oxygen ionization stages (right panel). In each panel, three characteristic points in the outflow are marked, including one below the Mach = 1 surface. In the bottom row, we additionally show a breakdown of the force multiplier by atomic species at each of these locations. 

Fig.~\ref{fig:loglog_bars} illustrates two key points. First, the characteristic force multiplier in the wind-launching region (near the Mach = 1 surface) is $\mathcal{M} \simeq \textrm{a~few} \times 100$. This is (just) high enough for a line-driven wind to be produced in this simulation. However, it is far from the maximum force multiplier of $\mathcal{M} \simeq 4400$ expected for near-optimal conditions \cite[e.g.][]{Proga1998, Higginbottom2024}, which {\em could} drive a strong wind in such a system. 

Second, the dominant driving species in all three of our test locations is oxygen (mainly O\,\textsc{iv}), followed by neon and nickel. This is quite different from the powerful line-driven winds of hot stars, where transitions associated with iron tend to dominate the line force \cite[e.g.][]{Vink1999, Noebauer2015}. This difference is a direct consequence of the different ionization states of -- and characteristic optical depths in -- these outflows. The relevant ionic species associated with Fe produce a huge number of weak lines, whereas those associated with O produce a smaller number of strong lines. High mass-loss rates tend to produce conditions in which strong lines are highly optically thick. Their contribution to the driving force then saturates, whereas that associated with weak lines can still grow in this regime. By contrast, the disc wind in our Model~A is too weak to be optically thick even to strong lines. Since they are not saturated in this regime, they dominate the total driving force.

\section{Discussion} \label{sec:Discussion}

In \cite{Higginbottom2024}, we presented the first radiation-hydrodynamic simulations of line-driven disc winds with a detailed multi-dimensional treatment of ionization and radiative transfer. Our main result was that -- at least in AWDs, but probably also in QSOs -- it is much more difficult to generate the empirically inferred high mass-loss rates via this mechanism than expected previously. The fundamental reason is the susceptibility of line-driven winds to over-ionization: when exposed to a strong ionizing radiation field, the ionization state of the flow can easily tip into a regime in which bound-bound and bound-free opacities are too low for efficient line driving. The main remaining limitation of these simulations was that the outflow was assumed to be isothermal.

In the new simulations presented here, we have been able to relax this isothermal approximation, while also improving our treatment of ionization and radiative transfer. We now explicitly and simultaneously solve for both ionization and thermal equilibrium throughout the flow. Despite these improvements, our results are very similar to those obtained in the isothermal approximation. Most importantly, over-ionization remains a problem, and the simulated outflows still produce much lower mass-loss rates ($\dot{M}_{\rm wind}/\dot{M}_{\rm acc} < 10^{-5}$; Table~\ref{tab:models}, also see Higginbottom et al. 2024\nocite{Higginbottom2024}) than earlier, more approximate simulations (which produced $\dot{M}_{\rm wind}/\dot{M}_{\rm acc} \simeq 10^{-4}$ for the same system parameters; see Proga et al. 1998 and Proga et al. 1999\nocite{Proga1998, proga1999}). They also produce much weaker spectroscopic signatures than seen in observations.

Possible resolutions to this discrepancy between the predictions of line-driven disk wind theory and observations were discussed and explored by \cite{Higginbottom2024}. For example, the over-ionization problem may be mitigated if the outflow is highly structured ("clumpy") on small, sub-grid scales or if the irradiating SED is softer than produced by a standard Shakura-Sunyaev disc. Additionally or alternatively, it is of course possible that mechanisms other than line-driving -- perhaps related to magnetic fields \cite[e.g.][]{Blandford1982,Scepi2019} -- may contribute to (or even dominate) the generation of these winds.

However, it is important to emphasize a critical point in this context. {\em The observed spectra of high-state AWDs and BAL QSOs show unambiguously that the outflows in these systems {\bf do} manage to avoid over-ionization.} As illustrated in Fig.~\ref{fig:loglog_bars} (for AWDs), they present precisely the strong UV resonance lines that are produced by the line-driven winds of hot stars. This suggests that line-driving {\em can} be significant in these systems, even if other physical processes may also have to be involved in launching the observed outflows.

While there is much work left to be done to understand whether/how line-driving works in AWDs, it is clearly also important now to investigate its relevance for AGN feedback. Line-driving is commonly assumed to be the mechanism responsible for powering the outflows from these systems, but no hydrodynamic simulations with a multi-dimensional treatment of ionization and radiative transfer have been carried out to date. In this context, our relaxation of the isothermal approximation is likely to be quite important, since the SEDs of AGN are more complex than those of high-state AWDs. It is difficult to predict the outcome of such simulations. On the one hand, AGN and QSOs can reach much higher Eddington ratios than AWDs, making it easier for radiative driving to work. On the other hand, over-ionization is also likely to be a significant challenge in these systems.

\section*{Acknowledgements}

We would like to thank the referee for their insightful comments and constructive suggestions, which have helped to enhance the clarity and quality of this manuscript. AM and CK were supported by the UK's Science \& Technology Facilities Council (STFC) grant ST/V001000/1. AW was supported by STFC studentship grant 2750006. Partial support for KSL's effort on the project was provided by NASA through grant numbers HST-GO-16489 and HST-GO-16659 and from the Space Telescope Science Institute, which is operated by AURA, Inc., under NASA contract NAS 5-26555. SAS acknowledges funding from STFC grant
ST/X00094X/1. JHM acknowledges funding from a Royal Society University Research Fellowship (URF\textbackslash R1\textbackslash221062).

\section*{Data Availability}

\textsc{Sirocco} is freely available on \href{https://github.com/sirocco-rt/sirocco}{GitHub}, with documentation accessible via \href{https://sirocco-rt.readthedocs.io/en/latest}{ReadTheDocs}. The specific version of the combined \textsc{Pluto-Sirocco} code used in this work (\textsc{Pluto-Sirocco} v1.0) is archived on \href{https://doi.org/10.5281/zenodo.15792686}{Zenodo (DOI:10.5281/zenodo.15792686)} and can also be found at \href{https://github.com/sirocco-rt/pluto-sirocco}{GitHub}. The simulations were performed using a modified version of \textsc{PLUTO} v4.4 (Patch 3), coupled with the \textsc{Sirocco} Monte Carlo radiative transfer code and an external CAK solver. Custom problem generators and input files are available from the authors upon request. Density movies of the simulation runs are provided as online supplementary material.



\bibliographystyle{mnras}






\bsp	
\label{lastpage}
\end{document}